\documentclass[conference]{IEEEtran}
\IEEEoverridecommandlockouts
\usepackage{amsmath,amssymb,amsfonts}
\usepackage{graphicx}
\usepackage{textcomp}
\usepackage{xcolor}
\usepackage[numbers,comma]{natbib}
\def\BibTeX{{\rm B\kern-.05em{\sc i\kern-.025em b}\kern-.08em
    T\kern-.1667em\lower.7ex\hbox{E}\kern-.125emX}}
\usepackage[nolist,nohyperlinks]{acronym}
\usepackage[T1]{fontenc}
\usepackage[utf8]{inputenc}
\usepackage[hyphens]{url}
\usepackage[hidelinks,breaklinks=true]{hyperref}
\usepackage{makecell}
\usepackage{multicol}
\usepackage{multirow}
\usepackage{todonotes}

\usepackage[nolist,nohyperlinks]{acronym}

\usepackage{comment}
\usepackage{enumitem}

\usepackage[ruled,vlined]{algorithm2e}
\SetKwComment{tcp}{// }{}
\SetKwProg{Fn}{Function}{}{}
\SetKwFunction{FHandleEvent}{HandleBudgetEvent}
\SetKwFunction{FAdjustBudget}{AdjustBudgetLease}
\SetKwFunction{FSelectNode}{SelectNodeWithBudget}
\SetKwFunction{FUpdateApp}{UpdateApplication}
\SetKw{KwAnd}{and}
\SetKw{KwOr}{or}
\SetKw{KwNot}{not}
\DontPrintSemicolon

\newcommand{\coo}{\ensuremath{\mathrm{CO_2}}}

\usepackage{tikz}
\usepackage{textcomp}

\newcommand\copyrighttext{%
  \footnotesize \textcopyright 2026 IEEE. Personal use of this material is permitted.  Permission from IEEE must be obtained for all other uses, in any current or future media, including reprinting/republishing this material for advertising or promotional purposes, creating new collective works, for resale or redistribution to servers or lists, or reuse of any copyrighted component of this work in other works. 

  Accepted as a paper at the 1st International Workshop of Software Architecture for Green Sustainable Carbon-Aware Software Systems Co-located with ICSA 2026.}
\newcommand{\copyrightnotice}{%
\begin{tikzpicture}[remember picture,overlay]
\node[anchor=south,yshift=10pt] at (current page.south) {\fbox{\parbox{\dimexpr\textwidth-\fboxsep-\fboxrule\relax}{\copyrighttext}}};
\end{tikzpicture}%
}

\begin{document}

\begin{acronym}
    \acro{ict}[ICT]{Information~and~Communications~Technology}
    
    \acro{euets}[EU ETS]{EUropean~Emissions~Trading~System}
    \acro{ets}[ETS]{Emissions~Trading~System}

    \acro{mapek}[MAPE-K]{Monitor, Analyze, Plan, Execute, with shared Knowledge}

    \acro{cu}[CU]{Computing~Units}

    \acro{rapl}[RAPL]{Running~Average~Power~Limit}
    \acro{nvml}[NVML]{NVIDIA~Management~Library}
    \acro{amdsmi}[AMD~SMI]{AMD~System~Management~Interface}
\end{acronym}

\title{Using Budgets to Reduce Application Emissions}

\def\orgsdu{SDU Software Engineering}
\def\sdu{University of Southern Denmark}
\def\locsdu{Odense, Denmark}

\author{

\IEEEauthorblockN{1\textsuperscript{st} Leo Wilhelm Lierse}
\IEEEauthorblockA{\textit{\orgsdu} \\
\textit{\sdu} \\
\locsdu \\
\href{mailto:leoli@mmmi.sdu.dk}{leoli@mmmi.sdu.dk}
}

\and

\IEEEauthorblockN{2\textsuperscript{nd} Mahyar Tourchi Moghaddam}
\IEEEauthorblockA{\textit{\orgsdu} \\
\textit{\sdu} \\
\locsdu \\
\href{mailto:mtmo@mmmi.sdu.dk}{mtmo@mmmi.sdu.dk}
}

\and

\IEEEauthorblockN{3\textsuperscript{rd} Sebastian Werner}
\IEEEauthorblockA{\textit{Information Systems Engineering} \\
\textit{Technische Universität Berlin} \\
Berlin, Germany \\
\href{mailto:werner@tu-berlin.de}{werner@tu-berlin.de}
}

}

\maketitle
\copyrightnotice

\begin{abstract}
As carbon pricing mechanisms like the EU Emissions Trading System are set to increase prices of energy consumption, software architects face growing pressure to design applications that operate within financially predictable emission constraints. 
Existing approaches typically enforce rigid per-interval emission rates, which prove unsuitable in electrical grids with highly dynamic carbon intensity, which is common in grids with growing renewable energy adoption.
We propose the use of \textit{emissions budgets}, an approach that replaces fixed emission rates with time-bound budgets, enabling applications to dynamically save unused emission allowances during low carbon intensity periods and expend them during high carbon intensity periods.
We describe emissions-aware applications using a MAPE-K feedback loop that continuously monitors application power consumption and grid carbon intensity, then adapts resource allocation through vertical scaling or migration to maintain long-term emission limits while maximizing performance.
Through simulation using six weeks of real-world carbon intensity data from Germany, France, and Poland, we demonstrate that budget-based management improves task fulfillment by up to 36\% in variable grids compared to fixed rates.
Crucially, budgets achieve parity with fixed rates in stable grids, making them a safe replacement. We show that emissions budgets are a practical mechanism to balance environmental constraints, operational costs, and service quality when emissions directly translate to financial penalties.
\end{abstract}
\begin{IEEEkeywords}
Emissions-Aware Computing, Sustainable Computing, Emissions Accounting, MAPE-K, Adaptive Resource Management
\end{IEEEkeywords}

\section{Introduction}\label{sec:introduction}

Climate change forces policymakers to act to reduce emissions across all parts of life and all sectors of the economy.
Carbon pricing mechanisms, such as emissions certificates or carbon taxes, are especially popular because they promote reductions in energy use, a shift to low-emission fuels, innovation, and the adoption of low-emission technologies, and generate revenues for implementing environmental or general policy \cite{black_carbon_2022}.
In most implementations, the costs will rise exponentially until the stated goal of zero emissions is met.
An example is the \ac{euets}\footnote{\url{https://climate.ec.europa.eu/eu-action/carbon-markets/about-eu-ets_en}}\footnote{\url{https://climate.ec.europa.eu/eu-action/carbon-markets/ets2-buildings-road-transport-and-additional-sectors_en}} which covers all emissions from power generation, over buildings to transportation.
In the short term, the price is expected to rise to over 100€~\cite{enerdata_carbon_2025}, \cite{pahle_eu-ets_2022} and in the long term (until 2050) up to over 700€ per ton of \coo-equivalent emissions~\cite{enerdata_carbon_2025}. 
We expect that these costs will flow downstream to application operators in the form of variable pricing based on the emissions of the total used energy.
This incentivizes active management of emissions to prevent volatile costs.

Existing approaches to curb application emissions often focus on shifting work to places (spatial shifting) \cite{sukprasert_limitations_2024}, \cite{asadov_carbon-aware_2025} or times (temporal shifting) \cite{wiesner_lets_2021}, \cite{sukprasert_limitations_2024}, \cite{asadov_carbon-aware_2025} where or when the used energy is less carbon-intensive, or they try to use only renewable energy \cite{wiesner_cucumber_2022}, \cite{wiesner_fedzero_2024}.
That, however, requires the application and the infrastructure on which it is hosted to support spatial/temporal shifting or renewable-only operation.
Another approach is to adjust how the application operates based on carbon intensity~\cite{thiede_carbon_2023}, \cite{wiesner_carbon-aware_2025}, i.e., there is a trade-off between service quality and energy efficiency.
These approaches minimize the total emissions, but are not appropriate for the described issue, where we need to hard limit our emissions to manage financial risks.

A fixed emissions rate is an instantaneous rate constraint (e.g., $gCO_2/s$) applied to an application.
This is enforced using vertical scaling or migrations, as in Carbon~Containers~\cite{thiede_carbon_2023}, for example, and is effective when the grid carbon intensity is stable.
That is usually the case in grids with lots of fossil (e.g., Poland) or non-intermittent green (e.g., France, Sweden) based power generation.
However, many countries in Europe currently operate a mixed grid with a high number of intermittent green energy sources (e.g., Solar, Wind) and have more dynamic carbon intensities.
Furthermore, there is an increasing number of datacenters that can operate partially on power from behind-the-meter installed capacity or have access to batteries \cite{colthorpe_google_2022}, which can further increase the dynamicity.
In such environments, fixed rates are unsuited: set too high, they underutilize the emission limit; set too low, they risk excessive throttling and suspension.
Right-sizing becomes infeasible.

An alternative approach to managing hard emissions limits is a budget, which allocates a cumulative emissions allowance over a defined lifetime.
Unlike fixed rates, budgets permit temporary emissions spikes during periods of high workload or carbon intensity, provided the cumulative allowance is respected over time.
Furthermore, budgets avoid the waste inherent with rates by preserving unused allowances for future use, and they support cost management in contexts where exceeding caps incurs penalties. 

In this paper, we intend to answer how emission budgets compare with fixed emission-rate limits in balancing the trade-off of total emissions, ETS-driven costs, and application performance.

\acused{mapek}

To do this, we frame emissions-aware self-adaptation using the MAPE-K (\acl{mapek}) architecture, a control loop comprising \textit{Monitor}, \textit{Analyze}, \textit{Plan}, and \textit{Execute} phases sharing a \textit{Knowledge} base~\cite{ibm_architectural_2005}.
While \citet{souza_ecovisor_2023} evaluated budgets for reducing emissions and limiting service quality degradation during concurrent high load and carbon intensity, we extend this work by:
(1) providing a general, MAPE-K architecture for self-adaptive emissions-aware applications, (2) comparing budget, fixed rate, and non-hard limit policies across a broad set of countries, and (3) looking at the financial impact of each policy.

The rest of the paper is structured as follows:
Section \ref{sec:background} introduces background and related work on emissions-aware computing.
Section \ref{sec:approach} presents our approach.
Section \ref{sec:implementation} provides details on the simulation we ran to test our approach.
Section \ref{sec:evaluation} reports the obtained results, and Section \ref{sec:discussion} discusses them.
Section \ref{sec:threats} discusses threats to validity.
Section \ref{sec:future} provides inspiration for future work.
Section \ref{sec:conclusion} concludes the paper.

\section{Background}\label{sec:background}

\subsection{Carbon Intensity}\label{sec:background.carbon-intensity}

The carbon intensity measures the emissions per kWh of energy produced.
As energy demand on a grid changes throughout the day, power plants come online or go offline.
This changes the mix of power sources, resulting in a change in carbon intensity.
If fossil-fuel-fired power plants come online, the carbon intensity increases.
If they get shut down or low-emissions power plants come online, the carbon intensity decreases.
Renewable energy introduces an additional source of variability because its generating capacity changes based on environmental factors.
Countries like Germany, with a significant amount of solar and wind power generating capacity, experience more pronounced swings throughout the day, depending on whether solar and wind power are available.
Missing renewable power generation means that fossil fuel-based reserve plants must be activated.

The variability of the carbon intensity in a grid can be described using a Coefficient of Variation\footnote{A CV of zero (0) indicates constant or nearly constant data, while a CV of over 1 means that the data varies more than the value of the mean.}~(CV)~\cite{thiede_carbon_2023}, the ratio of the standard deviation to the mean.
A greater use of renewable energy in a grid increases the variation of the carbon intensity.
Primarily because of the use of environmentally dependent energy sources.
Notable exceptions to that are hydro, geothermal, and nuclear energy, which produce low-carbon-intensity power without the high variability.

\subsection{Emissions Aware Applications}\label{sec:background.emissions-aware}

Emissions from applications can be reduced through two primary approaches: 
Improving energy efficiency, i.e., accomplishing the same work while consuming less energy.
For example, by using efficient infrastructure and hardware, using efficient algorithms, or utilizing existing resources more efficiently~\cite{jones_how_2018}, \cite{freitag_climate_2021}.
Or we improve emissions efficiency, i.e., using energy in a way that reduces emissions, for example, by sourcing cleaner power.
Focusing on emissions efficiency opens up a few additional strategies that incorporate information about the carbon intensity of the power used by the application.
We group those approaches into two main groups~\cite{green_software_foundation_carbon_nodate}: Demand Shifting and Demand Shaping.

\emph{Demand Shifting} comprises all approaches where we move applications through time or space.
For Temporal Shifting (shifting through time), we postpone work until the carbon intensity of the used energy is lower~\cite{wiesner_lets_2021}.
For example, in Germany, we can achieve great emissions reductions by postponing work until noon, when solar power reduces the carbon intensity of the grid.
Spatial Shifting (shifting through space) means we move our workload to another location \cite{sukprasert_limitations_2024}, \cite{asadov_carbon-aware_2025}.
For that, a location with low carbon intensity and low variability is selected, for example, France.
Executing the workload in that location inherently reduces emissions.
However, both temporal and spatial shifting usually require that the shifted workload or application actually supports that.
For temporal shifting, that means that work needs to support being postponed \cite{wiesner_lets_2021}.
For spatial shifting, latency, data locality, and data protection are concerns \cite{sukprasert_limitations_2024}.

\emph{Demand Shaping} includes all approaches that adjust the application based on carbon intensity instead of using temporal or spatial shifting.
That could be some form of shutting down parts of~\cite{wiesner_fedzero_2024}, adjusting service quality of~\cite{wiesner_carbon-aware_2025}, or throttling the application~\cite{thiede_carbon_2023}.
However, demand shaping approaches are usually applicable for a wider range of applications.
An example is an online store that shuts down recommendations when carbon intensity is high.
This reduces emissions, but impacts the service quality of the shaped application.

\subsection{Measuring Power Usage of Applications}

\acused{rapl}
\acused{nvml}
\acused{amdsmi}

One of the most important metrics for building energy- and emissions-aware applications is the power usage of an application.
However, accurately attributing power usage (and in turn energy usage) remains challenging.
When access to the hardware is available, for example, in on-premise deployments, then applications can get power telemetry through hardware interfaces.
For example, most mainstream CPUs are now providing access to power sensors that conform to the \ac{rapl}~(\acl{rapl}) interface.
These measurements can be correlated with processes (and thus applications) using performance counters or cgroup accounting to approximate power and energy usage.
Existing implementations like Kepler\footnote{\url{https://github.com/sustainable-computing-io/kepler}} or Scaphandre\footnote{\url{https://github.com/hubblo-org/scaphandre}} can provide this information as time series data.
In virtualized or some containerized settings, energy is estimated based on resource utilization metrics with the help of models \cite{colmant_process-level_2015}, \cite{zhang_estimating_2020}.
Some cloud providers also expose estimated energy consumption through APIs\footnote{\url{https://aws.amazon.com/sustainability/tools/aws-customer-carbon-footprint-tool/}}~\footnote{\url{https://learn.microsoft.com/en-us/industry/sustainability/}}~\footnote{\url{https://cloud.google.com/carbon-footprint}}, however, those are usually too coarse-grained and slow for actual use.

A key limitation of those mentioned approaches is that there is currently no way to actually account for the power usage for all components of a node (storage, fans, etc.), as the mentioned interfaces are component-specific.
Furthermore, they also exclude infrastructure energy usage, like cooling, networking, or storage (if separate).
And they also exclude other parts of the lifecycle, specifically the construction of the building that hosts the nodes and the manufacturing of the components.
Combined, these exclusions account for 21\% to 50\% of life cycle emissions~\cite{freitag_climate_2021}, \cite{gupta_chasing_2022}, \cite{lin_quantifying_2023}.

\subsection{Accounting for Sustainability}

For counting emissions, there are many different ways.
One of the simplest is to not count emissions at all, instead we group renewable energy sources into green energy and only use this green energy \cite{wiesner_cucumber_2022}, \cite{wiesner_fedzero_2024}.
Demand shifting-based approaches work especially well.
For example, a workload could be postponed until only solar energy is powering the grid, or at least the node.
This kind of accounting is also often linked with Smart-~and~Micro-grids.
Another concrete example is FedZero~\cite{wiesner_fedzero_2024}, which orchestrates federated ML training to only use nodes that have green energy available.
These approaches, however, rely on the availability of low-emissions energy sources locally, and don't make sense if those are not available (or measurable).

Another approach is to include batteries in the environment.
EcoVisor~\cite{souza_ecovisor_2023}, for example, enables software-defined batteries.
Those can be charged with the available energy, and the applications can switch between the grid and battery as a power source to improve emissions efficiency.
That, however, relies on this storage infrastructure actually being available.

Finally, on the emissions side, some approaches use either rates to limit emissions.
The benefit of those is that they do not depend on any infrastructure, like batteries or local power sources.
Carbon Containers \cite{thiede_carbon_2023}, for example, sets a fixed emissions rate under which emissions are kept.
Budgets, on the other hand, are an alternative approach to emissions accounting that allocates a cumulative emissions allowance over a defined lifetime~\cite{souza_ecovisor_2023}.
Unlike fixed rates, budgets permit temporary emissions spikes during periods of high workload or carbon intensity, provided the cumulative allowance is respected over time.
Furthermore, budgets avoid the waste inherent in fixed rates by preserving unused allowances for future use and supporting cost management in contexts where exceeding caps incurs penalties.
The budget lifetime should align with the shorter of the operator's billing cycle or the emissions pricing cycle.
Both rates and budgets use vertical scaling and migration to adjust emissions.

\section{Approach}\label{sec:approach}

We propose an architecture based on MAPE-K for self-adaptive emissions-aware applications.
We introduce an \emph{emissions budget} $B$ as an adaptation constraint that caps cumulative carbon emissions over a planning horizon~$T$.
Crucially, the architecture permits runtime deviation from the average allocation ($B/T$), enabling application-specific strategies to exploit variations in carbon intensity while guaranteeing hard emissions limits.
This empowers application architects to stabilize emissions costs without relying on generic adaptation policies.

\subsection{Architecture}

\begin{figure}[htb]
    \centering
    \vspace{0.2cm}
    \includegraphics[width=\columnwidth]{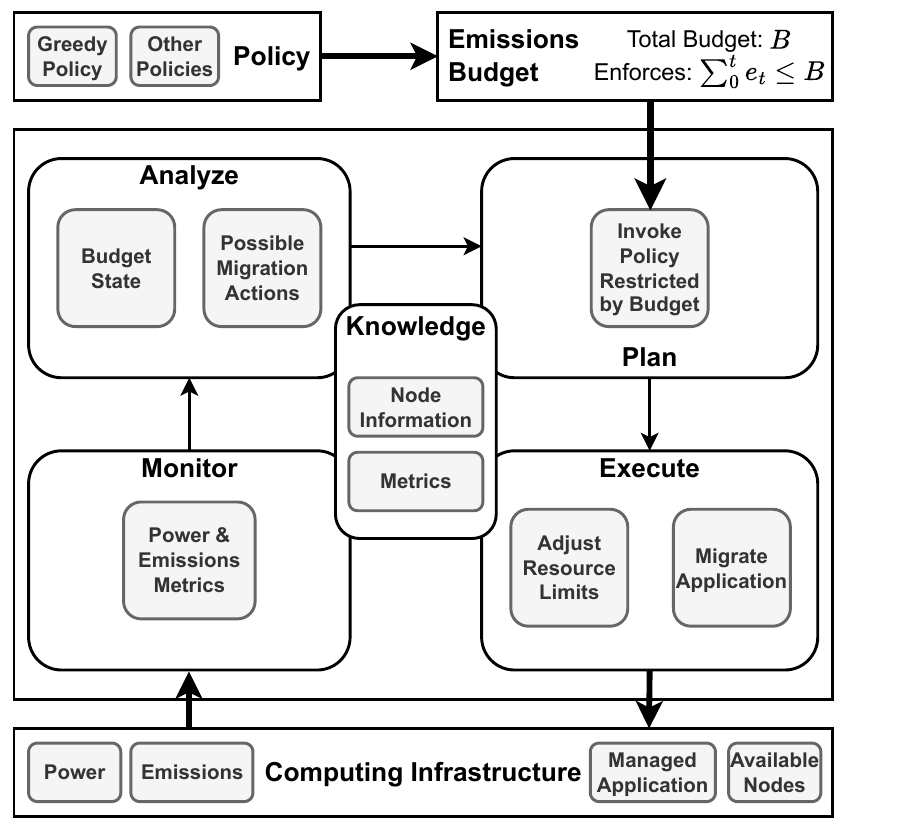}
    \caption{Emissions-Aware Self-Adaptation Architecture.}
    \label{fig:approach.architecture}
\end{figure}

Figure \ref{fig:approach.architecture} shows the architecture.
A self-adaptive application running on computing infrastructure (a cluster of nodes) is adapted through a MAPE-K loop.
In the \emph{Monitor} phase, the application's power consumption and the carbon intensity of the power supply are measured.

In the \emph{Analyze} phase three assessments are performed:
\emph{i)} the state of the budget $b_t$, including the remaining amount, and the current use rate.
\emph{ii)} possible migration targets $M_t$ are identified based on the power usage information available in the shared knowledge base, where migration is considered viable only if it reduces total energy consumption while maintaining or increasing available resources compared to current access~\cite{thiede_carbon_2023}.
\emph{iii)} and the current resource utilization $r_t$ is determined.

In the \textbf{Plan} phase, a policy $\pi$ selects an adaptation action $a \in \{ \text{scale up}, \text{scale down}, \text{migrate}, \text{no action} \}$ based on inputs from the knowledge base and the analysis results.
The policy's action space is constrained by the cumulative emissions budget $B$, which enforces $\sum_0^t e_t \leq B$.
Policies may operate freely until budget exhaustion restricts further carbon-intensive actions.
Critically, if the budget is exhausted before the budget's lifetime ends (and the budget is refilled), the application is suspended.
Applications using below their average budget allocation effectively "save" budget for future periods (e.g., to handle load or carbon intensity spikes), whereas over-consumption constitutes front-loading, i.e., "borrowing" against future allocations.
Thus, sustained overuse without subsequent reductions in carbon intensity or demand leads to tighter average allocation later in the budget's timeline.

Finally, in the \textit{Execute} phase, the selected action $a$ is enacted through vertical scaling or migration of the application.
The outcomes of execution are recorded in the knowledge base to inform subsequent control cycles.
Both the \textit{monitor} and \textit{execute} phases rely on the existing infrastructure to provide access to the required metrics or APIs for execution, respectively.

\subsection{Policies}
\label{sec:approach.policies}

A policy $\pi$ maps an application state, consisting of the emissions budget state $b_t$, feasible migration targets $M_t$, and current resource demands $r_t$, to concrete adaptation actions
\begin{equation*}
    \pi(b_t, M_t, r_t) \rightarrow \{ \text{scale up}, \text{scale down}, \text{migrate}, \text{no action} \}
\end{equation*}
Vertical scaling (scale up \& scale down) incurs negligible overhead but may be inefficient if the current node is less energy efficient under low load or scaling up hits a resource ceiling.
Migration enables access to nodes with different capacities and idle powers but incurs transfer overheads.

Policies must be developed application-specifically due to differing requirements and constraints between different applications and application types.
As an example, batch workloads are usually characterized as suspension-tolerant, whereas live services like APIs require uninterrupted availability, even if at degraded throughput or quality.

As a performance baseline for always-on applications (like a web API), we define a \textbf{greedy budget policy}: surplus budget is accumulated during low-demand periods and expended immediately when demand exceeds the baseline average allocation.
This approach prioritizes immediate responsiveness over long-term budget smoothing.
Crucially, the policy enforces a hard constraint; the average allocation never falls below the initial allocation, i.e., front-loading use of the budget is not allowed.
While this prevents premature budget exhaustion, it introduces another risk: during an extended period of high carbon intensity, the idle power draw of the host node may use the entire per‑second allocation.
Because the greedy policy spends all surplus immediately, there is no headroom to cover this while still performing useful work, which forces the application to be suspended.
A more forward-looking policy anticipates future emissions and does not spend surplus budget immediately, enabling it to operate for longer, even when carbon intensity is high.

Because it allocates the budget without looking ahead, this greedy policy uses the budget suboptimally.
Consequently, it establishes a performance floor: any adaptive policy achieving higher utility must demonstrably outperform this baseline.
It thus serves as a meaningful lower bound for evaluating more sophisticated budget-management techniques.

\section{Implementation}\label{sec:implementation}

To evaluate our approach, we implement a discrete-event simulation based on a modified version of the LEAF simulator~\cite{wiesner_leaf_2021}.
The implementation uses Python\footnote{\url{https://docs.python.org/3/}} and the SimPy framework\footnote{\url{https://simpy.readthedocs.io/en/latest/index.html}}.
We removed network modeling components and extended the simulator to track carbon emissions alongside power consumption.

\subsection{Simulation Setup}\label{sec:implementation.simulation-setup}

We model a heterogeneous cluster hosting a single application.
A baseline \textit{medium} node provides 100 \ac{cu} of capacity, with an idle power draw of 50W and a peak dynamic power draw of 600W under full utilization.
We derive \textit{small} and \textit{large} node configurations by halving and doubling these specifications, respectively.
Node power consumption follows a linear model based on utilization $U_{\text{node}}$:
\begin{equation} \label{eq:linear_power_model}
    P_{\text{node}} = P_{\text{idle}} + (P_{\text{peak}} - P_{\text{idle}}) \cdot U_{\text{node}}
\end{equation}
where $P_{\text{idle}}$ is the static idle power and $P_{\text{peak}}$ is the maximum power at 100\% utilization.
Since each node hosts exactly one application in our experiments, the application's power draw equals the node's total power consumption, intentionally including idle overhead.

Carbon emissions are computed per simulation second (our time resolution).
Given carbon intensity $\mathit{CI}$ in $\frac{\text{g}}{\text{kWh}}$, we convert to $\frac{\text{g}}{\text{Ws}}$ as:
\begin{equation} \label{eq:ci_conversion}
    \mathit{CI}_{\text{Ws}} = \frac{\mathit{CI}_{\text{kWh}}}{3,600,000}
\end{equation}
Emissions at time $t$ are then $e_t = P_{\text{node}} \cdot \mathit{CI}_{\text{Ws}}$.

\begin{figure}[ht]
    \centering
    \vspace{0.2cm}
    \includegraphics[width=0.95\columnwidth]{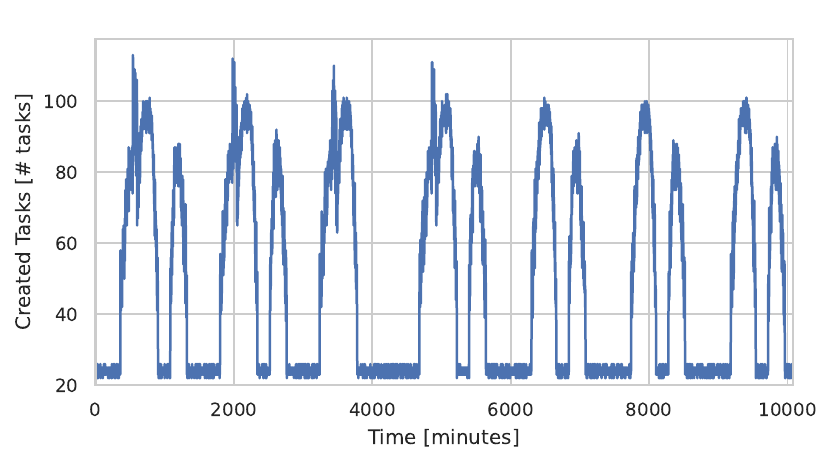}
    \caption{Synthetic task trace used for the experiments.}
    \label{fig:implementation.experiment-setup.task-trace}
\end{figure}
We simulate workloads using task objects, which arrive over time at the simulated application.
A task $t$ has a cu requirement $w_t$, a runtime $r_t$ and a deadline $D_t$.
The total required work is calculated as $W_t = w_t * r_t$.
A task is finished when sufficient \ac{cu} remain allocated over multiple simulation steps.
Figure~\ref{fig:implementation.experiment-setup.task-trace} shows our synthetic workload trace, which exhibits recurring diurnal patterns with two utilization peaks per day.

We do not simulate network latency or delays when scaling or migrating. 
Each application has perfect and instantaneous knowledge of the whole system.

\subsection{Experiment Setup \& Benchmarks}

\begin{figure}[ht]
    \centering
    \vspace{0.2cm}
    \includegraphics[width=0.95\columnwidth]{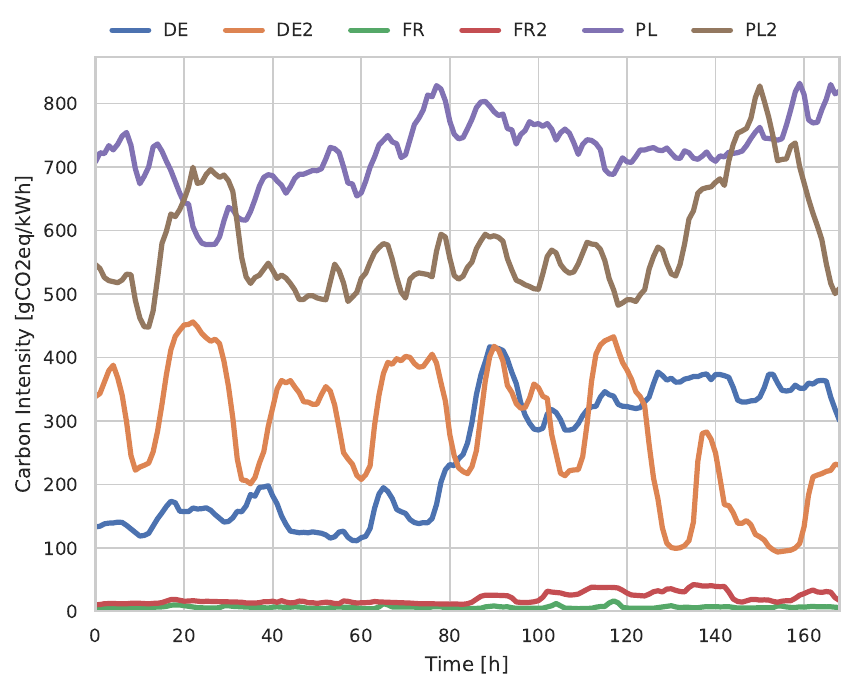}
    \caption{Carbon Intensity Data for the six selected weeks.}
    \label{fig:implementation.experiment-setup.carbon-intensity}
\end{figure}

We use historical carbon intensity data with an hourly resolution from Electricity~Maps\footnote{\url{https://www.electricitymaps.com/}}~(ODbL\footnote{\url{https://opendatacommons.org/licenses/odbl/1-0/}} licensed) for Germany~\cite{electricity_maps_germany_2025}, France~\cite{electricity_maps_france_2025}, and Poland~\cite{electricity_maps_poland_2025} in 2024.
After processing the data, we selected six one-week periods (two per country):
\begin{itemize}
    \item \emph{DE1}: 01.01.2024 to 07.01.2024
    \item \emph{DE2}: 22.04.2024 to 28.04.2024
    \item \emph{FR1}: 10.06.2024 to 16.06.2024
    \item \emph{FR2}: 23.12.2024 to 29.12.2024
    \item \emph{PL1}: 02.01.2024 to 08.01.2024
    \item \emph{PL2}: 20.01.2024 to 26.01.2024
\end{itemize}

Figure~\ref{fig:implementation.experiment-setup.carbon-intensity} shows the carbon intensity profiles.
France exhibits consistently low intensity with minimal variability, Poland shows high intensity with low to moderate variability, and Germany displays high variability.
This makes Germany the primary candidate for observing the benefits of using budgets instead of fixed limits.
We compare three execution policies:
\begin{enumerate}
    \item \textbf{Unlimited}: The emissions budget is ignored.
    It is not an actual policy that can be employed within the proposed architecture.
    However, it provides a useful comparison to observe possible efficiency losses due to having an emissions budget instead of "only" minimizing emissions.
    \item \textbf{Fixed Rate}: Constant execution with a $16.6mg/s$ emissions limit.
    \item \textbf{Greedy Budget}: Total budget of $10,080g$ allocated across the one-week runtime, equivalent to an average of $16.6mg/s$.
    Behaves as we described in Section \ref{sec:approach.policies}.
\end{enumerate}

We evaluate the policies by comparing their \textit{task fulfillment}, i.e., the total number of finished tasks, as the primary performance metric, to their power usage and emissions.

\section{Evaluation}\label{sec:evaluation}

\begin{table*}[th]
\centering
\caption{Experiment Results comparing all three strategies with 6 different carbon intensity profiles.}
\label{tab:evaluation.results}
\begin{tabular}{lc|cc|c|ccc|ccc|c}
                           &         & \multicolumn{2}{c}{Power {[}W{]}} & Energy {[}kWh{]} & \multicolumn{3}{c}{Emissions {[}mg{]}} & \multicolumn{3}{c}{Costs of Emissions {[}€{]}} & Finished Tasks {[}\#{]} \\
Scenario                   & Dataset & Mean            & Median          &                  & Mean     & Median   & \makecell{Total\\{[kg]}} & \makecell{80€\\{[/t \coo]}}  & \makecell{150€\\{[/t \coo]}}  & \makecell{700€\\{[/t \coo]}}  &                         \\ \hline
\multirow{6}{*}{Fixed}     & DE1     & 229.61          & 185.38          & 38.57            & 14.14    & 16.61    & 8.55             & 0.68 €       & 1.28 €         & 5.99 €         & 235962                  \\
                           & DE2     & 215.79          & 181.39          & 36.25            & 15.28    & 16.61    & 9.24             & 0.74 €       & 1.39 €         & 6.47 €         & 195951                  \\
                           & FR1     & 338.93          & 267.00          & 56.94            & 0.61     & 0.47     & 0.37             & 0.03 €       & 0.06 €         & 0.26 €         & 489862                  \\
                           & FR2     & 338.91          & 267.00          & 56.94            & 1.91     & 1.46     & 1.15             & 0.09 €       & 0.17 €         & 0.81 €         & 489864                  \\
                           & PL1     & 83.24           & 82.29           & 13.98            & 16.61    & 16.61    & 10.05            & 0.80 €       & 1.51 €         & 7.03 €         & 27196                   \\
                           & PL2     & 105.45          & 109.14          & 17.72            & 16.61    & 16.61    & 10.05            & 0.80 €       & 1.51 €         & 7.03 €         & 45971                   \\ \hline
\multirow{6}{*}{Budget}    & DE1     & 259.71          & 191.29          & 43.63            & 15.75    & 16.61    & 9.52             & 0.76 €       & 1.43 €         & 6.67 €         & 322257                  \\
                           & DE2     & 225.92          & 184.50          & 37.95            & 15.86    & 16.61    & 9.59             & 0.77 €       & 1.44 €         & 6.71 €         & 227560                  \\
                           & FR1     & 338.96          & 267.00          & 56.95            & 0.61     & 0.47     & 0.37             & 0.03 €       & 0.06 €         & 0.26 €         & 489857                  \\
                           & FR2     & 338.77          & 267.00          & 56.91            & 1.90     & 1.48     & 1.15             & 0.09 €       & 0.17 €         & 0.81 €         & 489857                  \\
                           & PL1     & 83.25           & 82.30           & 13.99            & 16.61    & 16.61    & 10.05            & 0.80 €       & 1.51 €         & 7.03 €         & 27207                   \\
                           & PL2     & 105.46          & 109.14          & 17.72            & 16.61    & 16.61    & 10.05            & 0.80 €       & 1.51 €         & 7.03 €         & 45980                   \\ \hline
\multirow{6}{*}{Unlimited} & DE1     & 338.92          & 267.00          & 56.94            & 23.18    & 17.89    & 14.02            & 1.12 €       & 2.10 €         & 9.82 €         & 489861                  \\
                           & DE2     & 338.93          & 267.00          & 56.94            & 25.11    & 21.25    & 15.19            & 1.22 €       & 2.28 €         & 10.63 €        & 489863                  \\
                           & FR1     & 338.93          & 267.00          & 56.94            & 0.61     & 0.47     & 0.37             & 0.03 €       & 0.06 €         & 0.26 €         & 489862                  \\
                           & FR2     & 338.91          & 267.00          & 56.94            & 1.91     & 1.46     & 1.15             & 0.09 €       & 0.17 €         & 0.81 €         & 489864                  \\
                           & PL1     & 338.92          & 267.00          & 56.94            & 67.70    & 53.42    & 40.95            & 3.28 €       & 6.14 €         & 28.66 €        & 489863                  \\
                           & PL2     & 338.94          & 267.00          & 56.94            & 53.74    & 42.21    & 32.50            & 2.60 €       & 4.88 €         & 22.75 €        & 489842                 
\end{tabular}
\end{table*}

Table~\ref{tab:evaluation.results} provides all results on power usage, energy usage, emissions, and task fulfillment. 

\subsection{Power Usage \& Task Fulfillment}
\label{sec:evaluation.power-and-tasks}

We first compare the power usage and task fulfillment between the three policies.
Both are closely related, as finishing tasks requires compute units, their use in turn uses power.
Firstly, we can confirm that the unlimited policy is working and has the same power usage and task fulfillment rate for all six experiments.
Comparing the fixed rate and greedy budget policies, we see that they have the same performance for both of the less variable locations.
In Poland, both throttle as much as possible.
And in France, both are not limited by the budget.
In Germany, we see differences for both DE1 and DE2; the greedy budget policy increases power usage and task fulfillment, while slightly increasing emissions.
For DE2, we observe a 5\% increase in energy usage and a 16\% increase in finished tasks.
The greedy budget policy achieved even larger improvements for DE1, with a 13\% increase in power usage and a 36\% improvement in task fulfillment.
\begin{figure}[ht]
    \centering
    \vspace{0.2cm}
    \includegraphics[width=0.95\columnwidth]{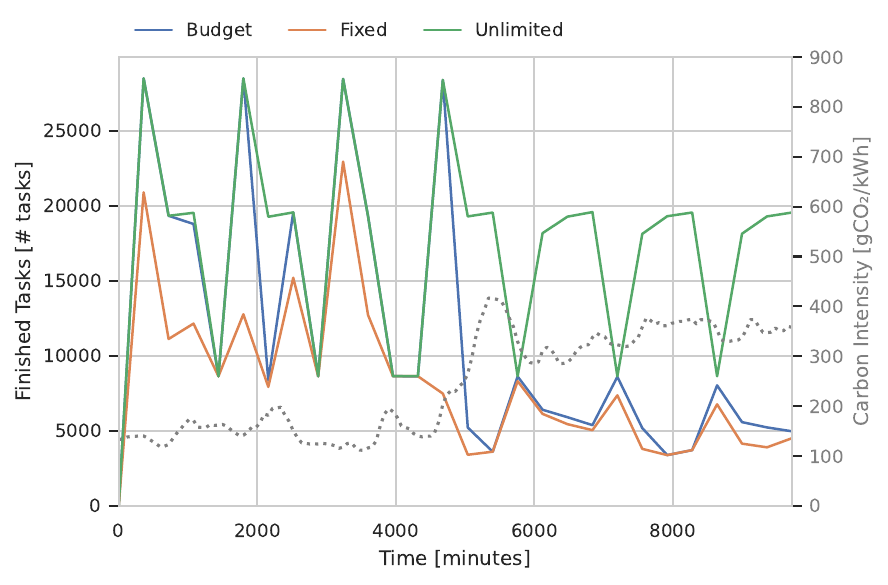}
    \caption{Comparison of the total number of completed tasks across strategies during the DE1 carbon-intensity period, aggregated as sums over six-hour intervals.}
    \label{fig:approach.power-and-tasks.finished-tasks-de1}
\end{figure}
Figure~\ref{fig:approach.power-and-tasks.finished-tasks-de1} visualizes the improvement of the greedy budget policy very well.
Between spikes, it is able to save some of the emissions budget and thus can compensate somewhat for them.
DE1 has a rise in emissions after the first four days, and thus, both the greedy budget and the fixed rate limit policies have the same task fulfillment rates after that.
Compared to the unlimited policy, the greedy budget policy reduces both power consumption and task fulfillment.
For the DE scenarios, the greedy budget policy consumes 23\% less energy for DE1 and 33\% less for DE2, while fulfilling 34\% fewer tasks for DE1 and 54\% fewer for DE2.
For the PL scenarios, which have higher emissions on average, the reductions are more pronounced: energy usage drops by 75\% for PL1 and 68\% for PL2, and task fulfillment falls by 94\% for PL1 and 90\% for PL2.

\subsection{Emissions}\label{sec:evaluation.emissions}

We compare the emissions between the three policies.
Keeping emissions below a target value is the requirement for both the fixed and the greedy budget policy.
Firstly, the unlimited policy is emitting by far the most, which is expected.
However, for FR1 and FR2, the unlimited policy has equally low emissions as the other two, as the carbon intensity of the French grid is very low.
Comparing the fixed rate and greedy budget policies, we again see that they have the same emissions for both France and Poland.
For both German carbon intensity periods, we see differences.
The greedy budget policy increases emissions by 11\% in DE1 and by 4\% in DE2, compared to the fixed rate policy.
\begin{figure}[ht]
    \centering
    \vspace{0.2cm}
    \includegraphics[width=0.95\columnwidth]{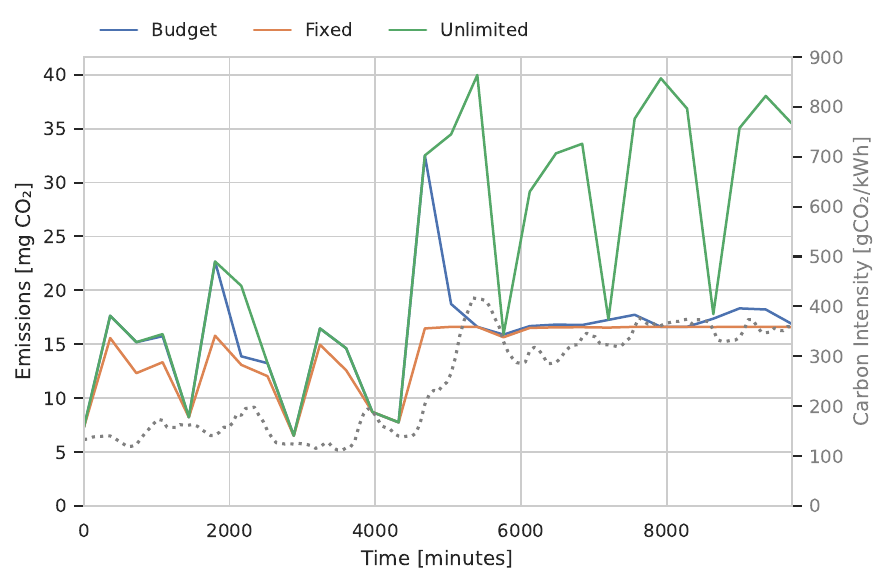}
    \caption{Comparison of emissions across policies during the DE1 carbon intensity period, aggregated as means across six-hour intervals.}
    \label{fig:approach.emissions.emissions-de1}
\end{figure}
Figure~\ref{fig:approach.emissions.emissions-de1} shows the emissions for all three policies using DE1.
During low-carbon-intensity periods, the greedy budget policy can save up and then compensate for load spikes.
The fixed rate strategy is unable to do that and thus stays fixed below its budget.
This is clearly visible in the first half of both Figures~\ref{fig:approach.power-and-tasks.finished-tasks-de1} and~\ref{fig:approach.emissions.emissions-de1}.
Compared to the unlimited policy, the greedy budget policy reduces total emissions by 32\% for DE1, 36\% for DE2, 75\% for PL1, and 69\% for PL2.

\subsection{Cost}\label{sec:evaluation.cost}

We calculate the cost of the emissions of the experiments using a carbon price of 700€ per ton of \coo equivalent emissions.
Over the experiment runtime, the applications used at most 350W, resulting in modest absolute costs.
Under the unlimited policy applied for PL1, we recorded the highest emissions, yielding the highest observed cost of 28.66€.
In contrast, both the fixed rate and greedy budget policies maintained costs below 7.10€ across all workloads, demonstrating their effectiveness in limiting cost.

\section{Discussion}\label{sec:discussion}

We argue that the main benefit of using budget-based policies is that they outperform fixed-rate policies in grids with variable carbon intensity.
In Section~\ref{sec:evaluation}, we show that using the greedy budget policy improves task fulfillment by up to 36\% in Germany with its highly dynamic grid, while increasing total emissions by only 13\% compared to the fixed rate policy, while still staying within the hard limit.
This is in line with results from \citet{souza_ecovisor_2023}, which demonstrated that application-specific policies using budgets perform better than fixed rate policies.
The modest emissions increase is expected as budgets target long-term average emissions rather than enforcing rigid short-term limits.
The relatively greater improvement in task fulfillment is due to the greedy budget policy being able to utilize the high-performance large node to process tasks during load spikes.
For countries with less variable grids, like France (low intensity) and Poland (high intensity), the greedy budget policy achieves parity with the fixed rate policy in both emissions and performance, confirming that budgets can safely replace fixed rates without degradation.

However, compared to the unlimited strategy, budgets face a fundamental limitation: they fulfill 34\% to 94\% fewer tasks while reducing emissions by only 32\% to 75\%.
This stems from idle power overhead; for example, our model allocates 50W static power per medium node.
At 54\% utilization, the medium node uses 350W, and static power makes up 14\% of the total power usage.
At 10\% utilization, it uses 105W, and static power makes up 48\% of the total power usage.
This explains why hard emissions-limited policies achieve less relative emissions reduction compared to the performance loss.
Consequently, an unconstrained system that fully utilizes resources achieves better energy efficiency and often better absolute emissions efficiency, though without hard emission guarantees.

Although absolute carbon costs remain modest during our single application experiments ($\leq$28.66€ using the unlimited policy), they compound linearly with infrastructure scale and operational duration.
Extrapolating to a 100-application system, the cost difference between unlimited and hard-limited policies is approximately 2,900€.
If the costs of emissions are passed on, staying within a hard limit or not can have a large impact on deployment costs.

In practice, practitioners will need to decide if they are fine with dynamic emissions-related costs.
If not, and a hard emissions limit is required to control those, practitioners should integrate an emissions budget-based policy into their system to limit emissions, instead of utilizing a fixed rate limit.
That said, any hard emissions constraint will be less energy- and emissions efficient than an unconstrained system, which is the trade-off of enforcing such an emissions limit.
The loss in performance may outweigh the benefits of predictable costs and sustainable operation, forcing practitioners to prioritize quality of service.

\section{Threats To Validity}\label{sec:threats}

Our reliance on simulation, which abstracts away real-world system complexities, may be a threat to the validity of the results.
Policies have instantaneous, noise-free access to the complete cluster state and react within the same simulation step to carbon intensity changes, enabling a very fine-grained adaptation.
Real systems face monitoring latency, communication delays, and control loop overheads that would dampen responsiveness and reduce the magnitude of improvements observed.
Similarly, application migrations occur instantaneously in our model; real migrations incur downtime, network transfer delays, and CPU overhead that would make aggressive migration strategies less attractive.
Carbon Containers~\cite{thiede_carbon_2023} mitigates this in practice through migration preemption policies to avoid thrashing.

Another threat stems from our power model's simplification of node power usage.
We assume a fixed idle-to-peak power ratio across all workloads and hardware configurations.
In reality, power usage characteristics of the nodes can vary based on the installed hardware~\cite{blem_power_2013} and dynamic power usage can behave non-linearly~\cite{khokhriakov_multicore_2020}. 

Finally, Our evaluation spans only three European grids, over one-week periods in 2024.
Grids in other regions may exhibit different characteristics that could affect the efficacy of budget-based policies.
Furthermore, our workload model represents a single application that handles time-sensitive requests.
Latency-sensitive or other kinds of services may exhibit different trade-offs between emissions constraints and performance degradation.
Finally, we evaluate at the application level; in production environments, interactions between co-located applications, shared infrastructure resources, and cluster-level scheduling policies could alter the observed emissions, performance, and energy trade-offs.

\section{Future Work}\label{sec:future}

In the future, we plan to implement emissions budgets in a production container orchestration system (e.g., Kubernetes) to assess their efficacy under real-world constraints such as monitoring latency, migration overheads, and multi-tenant interference.
Extending budgets to the infrastructure level would allow cluster managers to allocate carbon budgets across applications while balancing application requirements and facility-wide emission targets, enabling coordinated data center management.
Hybrid approaches that combine budget-based throttling with workload shifting or geographical migration could further optimize the emissions-performance trade-off.
Finally, more sophisticated policies, such as moderate use of excess budget, carbon intensity forecasting, and strategic front-loading of budget use, could improve both emissions efficiency and performance.

\section{Conclusion}
\label{sec:conclusion}

We demonstrate that emissions budgets offer a flexible and effective alternative to fixed rates for enforcing hard emissions limits.
Evaluated across diverse grid conditions, we show that budgets significantly improve task fulfillment by up to 36\% in variable grids, while maintaining long-term emission limits, and achieve parity with fixed limits in stable grids without performance degradation.
However, compared with applications that do not have a fixed long-term emission limit, they only reduce emissions by at most 75\%, while incurring a performance penalty of up to 94\%. 
This highlights a fundamental trade-off: 
Operating under emissions budgets achieves the highest emissions efficiency when hard emission limits are required.
But that comes at the cost of energy efficiency due to idle power overhead.
Thus, unconstrained systems that maximize resource utilization are more energy- and emissions-efficient overall but are not guaranteed to stay within emission limits.
As emissions pricing mechanisms reduce allowances or increase prices, emissions costs have an increased impact.
Budget-based emissions management provides a practical mechanism for applications to operate within predictable emission constraints while adapting to the dynamic carbon intensity of electrical grids with diverse energy sources.

{
\small
\bibliographystyle{IEEEtranN}
\bibliography{refs,electricitymaps}
}

\end{document}